\documentclass[a4paper,12pt]{article}

\usepackage{amsmath,amssymb}

\newcommand{{\dotprod } }{\stackrel{\scriptscriptstyle\bullet}{{}}}

\makeatletter
\renewcommand{\section}{%
  \@startsection{section}%
   {1}%
   {\z@}%
   {-3.5ex \@plus -1ex \@minus -.2ex}%
   {2.3ex \@plus.2ex}%
   {\normalfont\large\bfseries}%
}%
\makeatother

\makeatletter
\renewcommand{\subsection}{%
  \@startsection{subsection}%
   {1}%
   {\z@}%
   {-3.5ex \@plus -1ex \@minus -.2ex}%
   {2.3ex \@plus.2ex}%
   {\normalfont\large}%
}%
\makeatother

\begin{document}

\begin{center}
{\Large\bf Considering Relativistic Symmetry as the First Principle of Quantum Mechanics}
\\ \ \\
{\large Takuya Kawahara \footnote[1]{k\_takuya@st.rim.or.jp}} 
\\ \ \\
\textit{GAIA System Solutions Inc., Tokyo 141-0022, Japan}
\end{center}

\noindent
\textbf{Abstract:} \ On the basis of the relativistic symmetry of Minkowski space, 
we derive a Lorentz invariant equation for a spread electron. 
This equation slightly differs from the Dirac equation and includes 
additional terms originating from the spread of an electron. 
Further, we calculate the anomalous magnetic moment based on these terms. 
These calculations do not include any divergence; 
therefore, renormalization procedures are unnecessary. 
In addition, the relativistic symmetry existing among coordinate systems 
will provide a new prospect for the foundations of quantum mechanics like 
the measurement process. \\ 

\\
\noindent
\textit{PACS(2006): \ 03.65.Pm, 13.40.Em, 03.65.Ta, 03.65.Ud}

\section{\label{sec:level1}Introduction}
\hspace*{0.6cm}There are many problems associated with a relativistic 
quantum field theory. 
In particular, the issue of infinity accompanied by radiative correction 
is troublesome. 
Renormalization methods allow most of the divergence to be eliminated; 
however, it is difficult to accept this method as the final solution. 
In addition, much work has been done on the study of Dirac particles 
\cite{spohn}. 
Nevertheless, even today, quantum field theory continues to be problematic 
with regard to its relationship with the theory of relativity. 
Therefore, apart from the conventional approach, we will directly derive a 
Lorentz invariant equation for an electron on the basis of the symmetry of 
Minkowski space. 
Thus, we assume two fundamental principles, instead of the usual rules of 
quantum mechanics, as follows: 

\hangindent=2\parindent
(\,i\,) An electron has an inherent relativistic symmetry, i.e., the 
behavior of an electron is described as the function of only an invariant 
parameter in Minkowski space. 

\hangindent=2\parindent
(ii) An electron has a finite size as the world length that is proportional 
to the reciprocal of the inertial mass. 

\noindent
These principles imply that an electron identifies the Minkowski space as 
one-dimensional; this must be a cause of the quantum behavior of an electron. 

In Sec.2, we extract a relativistic invariant parameter in Minkowski 
space. 
In Sec.3, based on these principles, we derive a relativistic equation 
for a spread electron. 
This equation slightly differs from the Dirac equation and includes 
additional terms originating from the spread of an electron. 
These terms are interpreted as an enhanced Pauli term, which is related to 
the anomalous magnetic moment \cite{weinberg}. 
Up to now, the Pauli term has been disregarded because it makes 
renormalization impossible. 
Nevertheless, in Sec.4, we calculate the corrections in magnetic moment 
based on these terms, without renormalization. 
In addition, in Sec.5, the measurement process in quantum theory is 
discussed based on the relativistic symmetry existing among coordinate 
systems. 
\section{\label{sec:level2}Extraction of a Relativistic Invariant 
Parameter}
\hspace*{0.6cm}In this paper, we use Einstein's summation convention 
for indices 
$ {\mu },{\nu }, $ and $ {\xi } \; ( = 0, 1, 2, 3 ) $. 
Using the Minkowski metric 
$ \textit{\textsf{g}}_{\mu \nu } \! = diag(+,-,-,-) $, 
we define the relation 
between a covariant vector $ a_{\mu } $ and a contravariant vector 
$ a^{\nu } $ as follows: 
\begin{equation}
 a_{\mu } = \textit{\textsf{g}}_{\mu \nu } a^{\nu } . \label{201}
\end{equation}
Further, we substitute $ \hbar = c = 1 $ as a rule. 

According to the special theory of relativity, world length squared 
$ \delta s ^{2 } $ is a Lorentz invariant. 
For any inertial coordinate system, the following identity holds 
between a world length $ \delta s $ 
and coordinate intervals $ \delta x ^{\mu } $: 
\begin{equation}
 \delta s^{2 } = \textit{\textsf{g}}_{\mu \nu } \delta x^{\mu } 
 \delta x^{\nu } , 
 \ \ \ \mbox{where} \ \quad x^{\mu } 
 \equiv ( x^{0 } , x^{1 } , x^{2 } , x^{3 }) 
 \equiv ( t , \boldsymbol{x } ) .  \label{202}
\end{equation}
For convenience, we designate the origin of the inertial coordinate system as 
$ s = 0 $; thus, the quadratic form (\ref{202}) becomes 
\begin{equation}
 s^2 = \textit{\textsf{g}}_{\mu \nu } x^{\mu } x^{\nu } . \label{203} 
\end{equation}

In order to extract the relativistic invariant parameter $ s $, 
we take the square root of (\ref{203}) in the linear form. 
Now, we assume that the quadratic form (\ref{203}) is decomposed as follows: 
\begin{equation}
 s = \gamma _{\mu } x^{\mu } . \label{204}
\end{equation}
Let us square both sides of Eq.(\ref{204}):
\begin{equation}
 {s }^{2 } \! \! = \! \! \sum_{ {\mu } > {\nu } } 
 ( \gamma _{\mu } \gamma _{\nu } 
 + \gamma _{\nu } \gamma _{\mu } ) x^{\mu } x^{\nu } 
 \! \! + \! \frac{1 }{2 } \! \sum_{ {\mu } = {\nu } } 
 ( \gamma _{\mu } \gamma _{\nu } 
 + \gamma _{\nu } \gamma _{\mu } ) x^{\mu } x^{\nu } \! . \label{205}
\end{equation}
On the other hand, the quadratic form (\ref{203}) can be rewritten as follows: 
\begin{equation}
 s^2 = \sum_{ {\mu } > {\nu } } 2 \textit{\textsf{g}}_{\mu \nu } 
 x^{\mu } x^{\nu } 
 + \frac{1 }{2 } \sum_{ {\mu } = {\nu } } 2 \textit{\textsf{g}}_{\mu \nu } 
 x^{\mu } x^{\nu } . \label{206}
\end{equation}
Expressions (\ref{205}) and (\ref{206}) are equivalent when $ \gamma _{\mu } $ 
satisfies the following relation: 
\begin{equation}
 \gamma _{\mu } \gamma _{\nu } + \gamma _{\nu } \gamma _{\mu } 
 = 2 \textit{\textsf{g}}_{\mu \nu } . \label{207}
\end{equation}
It implies that $ \gamma _{\mu } $ are isomorphic forms of the 
Dirac $ \gamma $-matrices. 

We now introduce the rule of raising and lowering indices of 
the $ \gamma $-matrices as well as the vectors: 
\begin{equation}
 \gamma ^{0 } = \gamma _{0 } , \ \gamma ^{k } = - \gamma _{k } 
 \quad ( k = 1, 2, 3 ) . \label{208}
\end{equation}

Similar to Eq.(\ref{204}), the following identical equation holds: 
\begin{equation}
 \frac{d }{ds } \Psi = \gamma ^{\mu } \:\! \partial _{\mu } 
 \Psi , \  
 \mbox{where} \ \ \ 
 \partial _{\mu } \equiv \frac{\partial }{\partial x^{\mu } } 
 \equiv \Bigl( \frac{\partial }{\partial x^{0 } }, \frac{\partial }
 {\partial x^{1 } }, 
 \frac{\partial }{\partial x^{2 } }, \frac{\partial }{\partial x^{3 } } \Bigr) 
 \equiv \Bigl( \frac{\partial}{\partial t}, \boldsymbol{\nabla } \Bigr) . 
 \label{209} 
\end{equation}
When $ \Psi $ is a function only of argument $ s $, 
\begin{equation}
 \frac{d \Psi (s ) }{d s } = \frac{\partial x^{\mu } }{\partial s } \, 
 \frac{\partial \Psi (s ) }{\partial x^{\mu } } 
 = \gamma ^{\mu } \:\! \partial _{\mu } \Psi , \
 \mbox{where} \ \ \ \frac{\partial x^{\mu } }{\partial s } 
 = \Bigl( \frac{\partial s }{\partial x^{\mu } } \Bigr)^{-1 } 
 = \gamma _{\mu }^{-1 } = \gamma ^{\mu } . 
 \label{210}
\end{equation}
In this case, Eq.(\ref{209}) is valid. 
However, since the identical equation (\ref{209}) contains $ \gamma $-matrices 
in the operator, we consider $ \Psi $ as a four-component function. 
\section{\label{sec:level3}Derivation of Equations for a Spread Electron}
\hspace*{0.6cm}In this section, we derive a relativistic difference equation 
for a spread electron. 
Further, we derive a wave equation from this difference equation. 
The wave equation agrees with the Dirac equation except for the existence of 
additional terms originating from the spread of an electron. 
\subsection{\label{subsec:level31}Derivation of the difference equation}
Based on the principles referred to in Sec.1, we identify the space-time 
behavior of an electron using the following equation: 
\begin{equation}
 \rho ( s + \delta s ) = \rho (s ) , \label{301}
\end{equation}
where $ \rho $ denotes the density scalar and $ \delta s $ 
denotes the world length of the electron. 
This equation implies the conservation law for the existing probability of 
a spread electron under a proper time evolution. 
We transform Eq.(\ref{301}) into an equation for any inertial coordinate 
system as follows: 

We introduce an adjoint of $ \Psi $, 
$ \overline{\Psi } \equiv \Psi ^{\dagger } \gamma ^{0 } $, 
and define the density scalar $ \rho $ as follows: 
\begin{equation}
 \rho (s ) \; {\equiv } \; \overline{\Psi } (s ) \Psi (s ) . \label{302}
\end{equation}
Here, $ \Psi $ is a wave function of an electron, which will be clarified 
later. 
From definition (\ref{302}), $ \overline{\Psi } \Psi $, and not $ \Psi $, 
evidently relates to the existing probability of an electron. 
By substituting (\ref{302}) in Eq.(\ref{301}), we obtain 
\begin{equation}
 \overline{\Psi } ( s + \delta s ) \Psi ( s + \delta s ) 
 = \overline{\Psi } (s ) \Psi (s ) . \label{303}
\end{equation}
$ \Psi (s + \delta s ) $ is expressed as the Maclaurin series: 
\begin{equation}
 \Psi ( s + \delta s ) \to \exp \Big( \delta s \frac{d }{ds } \Big) 
 \Psi (s ) . \label{304}
\end{equation}
Since $ \Psi $ depends only on $ s $, (\ref{209}) can be substituted in 
(\ref{304}): 
\begin{equation}
 \exp \Big( \delta s \frac{d }{ds } \Big) \Psi (s ) 
 \to \exp ( \delta s \, \gamma ^{\mu } \:\! {\partial }_{\mu } ) 
 \Psi . \label{305}
\end{equation}
Similar to Eqs.(\ref{304}) and (\ref{305}), using the relation 
$ \gamma ^{0 } ( \gamma ^{\mu } )^{\dagger } \gamma ^{0 } = \gamma ^{\mu } $, 
we obtain 
\begin{eqnarray}
 \begin{aligned}
 \overline{\Psi } ( s + \delta s ) 
 \equiv \Psi^{\dagger } ( s + \delta s ) \gamma ^{0 } 
 \to \overline{\Psi } \, 
 \exp ( \delta s \, \gamma ^{\mu } 
 \overleftarrow{{\partial } }_{ \!\! \mu }) , 
 \hspace{5mm} \label{306} \\
 \mbox{where} \quad \overline{\Psi } \gamma ^{\mu } 
 \overleftarrow{{\partial } }_{ \!\! \mu } \; {\equiv } \; {\partial }_{\mu } 
 \overline{\Psi } \gamma ^{\mu } .  
 \end{aligned}
\end{eqnarray}

By introducing the $ 4 \times 4 $ square matrix $ U $ of parameter 
$ \delta s $, we assume the following equation: 
\begin{equation}
 \exp ( \delta s \, \gamma ^{\mu } \:\! {\partial }_{\mu } ) \Psi 
 = U ( \delta s ) \Psi . \label{307}
\end{equation}
Then, an adjoint equation of Eq.(\ref{307}) is expressed as follows: 
\begin{equation}
 \overline{\Psi } \exp ( \delta s \, \gamma ^{\mu } 
 \overleftarrow{{\partial } }_{ \!\! \mu } ) 
 = \overline{\Psi } \gamma ^{0 } U^{\dagger } ( \delta s ) \gamma ^{0 } . 
 \label{308}
\end{equation}
Multiplying each side of Eq.(\ref{308}) from the left with the corresponding 
sides of Eq.(\ref{307}), we obtain the following expression: 
\begin{equation}
 \overline{\Psi } \exp ( \delta s \, \gamma ^{\nu } 
 \overleftarrow{{\partial } }_{ \!\! \nu } ) 
 \exp ( \delta s \, \gamma ^{\mu } \:\! {\partial }_{\mu } ) \Psi 
 = \overline{\Psi } \gamma ^{0 } U^{\dagger } ( \delta s ) \gamma ^{0 } 
 U ( \delta s ) \Psi . \label{309}
\end{equation}
Therefore, matrix $ U $ should satisfy the relation: 
\begin{equation}
 \gamma ^{0 } U^{\dagger } ( \delta s ) \gamma ^{0 } U ( \delta s ) 
 = {\rm I } , \label{310}
\end{equation}
such that Eq.(\ref{309}) is equivalent to Eq.(\ref{303}). 
Here, we assume that matrix $ U (\delta s ) $ is continuous for 
$ \delta s $ and is expressed as follows by the introduction of the 
$ 4 \times 4 $ square matrix $ M $: 
\begin{subequations}
\begin{eqnarray}
 U \! & = & \! \exp ( {\rm i } \, \delta s M ) \label{311a} \\
 \mbox{and} \hspace{10mm} U^{\dagger } \! \! \! & = & \! 
 \exp ( - {\rm i } \, \delta s M^{\dagger } ) . \hspace{15mm} 
\label{311b}
\end{eqnarray}
\end{subequations}
Matrix $ U^{\dagger } $ can be expressed as 
\begin{equation}
 U^{\dagger } ( \delta \;\!\! s ) \! = \exp ( - {\rm i } \, \delta 
 \;\!\! s M^{\dagger } ) 
 = \lim _{n \to \infty } \Bigl( {\rm I } - \frac{ {\rm i } \, 
 \delta \;\!\! s } {n} M^{\dagger } \Bigr)^n . \label{313} 
\end{equation}
If matrix $ M $ is chosen to satisfy the relation: 
\begin{equation}
 \gamma ^{0 } M^{\dagger } \gamma ^{0 } = M , \label{312}
\end{equation}
we get 
\begin{eqnarray}
 \begin{aligned}
 \gamma ^{0 } U^{\dagger } \gamma ^{0 } U \! 
 = \lim _{n \to \infty } 
 \gamma ^{0 } \Bigl( {\rm I } - \frac{ {\rm i } \, \delta \;\!\! s }{n} 
 M^{\dagger } \Bigr)^n \gamma ^{0 } U \hspace{30mm} \\ 
 = \gamma ^{0 } \Bigl( {\rm I } - \frac{ {\rm i } \, \delta \;\!\! s } 
 {n} M^{\dagger } \Bigr) \gamma ^{0 } 
 \gamma ^{0 } \Bigl( {\rm I } - \frac{ {\rm i } \, \delta \;\!\! s } 
 {n} M^{\dagger } \Bigr) \gamma ^{0 } \cdots U \hspace{2.5mm} \\ 
 = \Bigl( {\rm I } - \frac{ {\rm i } \, \delta \;\!\! s }{n} M \Bigr) 
 \Bigl( {\rm I } - \frac{ {\rm i } \, \delta \;\!\! s }{n} M \Bigr) 
 \cdots U \hspace{21.5mm} \\
 = \lim _{n \to \infty } 
 \Bigl( {\rm I } - \frac{ {\rm i } \, \delta \;\!\! s }{n} M \Bigr)^n 
 = \exp ( - {\rm i } \, \delta s M ) \, U \hspace{11mm} \\ 
 = U^{-1} U = {\rm I } , \hspace{57mm} \label{314} 
 \end{aligned}
\end{eqnarray}
then matrix $ U $ satisfies (\ref{310}). 

Now, operating $ \gamma ^{0 } $ from the left-hand side of (\ref{312}), 
we have 
\begin{equation}
 M^{\dagger } \gamma ^{0 } = \gamma ^{0 } M . \label{315}
\end{equation}
On the other hand, 
\begin{equation}
 M^{\dagger } \gamma ^{0 } = M^{\dagger } ( \gamma ^{0 } )^{\dagger } 
 = ( \gamma ^{0 } M )^{\dagger } . \label{316}
\end{equation}
Hence, relation (\ref{312}) is equivalent to the condition that 
$ \gamma ^{0 } M $ is hermitian. 
Since $ \gamma ^{0 } \gamma ^{\mu } $ is hermitian, we adopt a linear 
combination of $ \gamma ^{\mu } $ as $ M $: 
\begin{equation}
 M = - {\rm e } \gamma ^{\mu } A_{\mu } , \label{317}
\end{equation}
where $ {\rm e } $ is the electron charge 
and $ {A }_{\mu } $ represents the four-vector potential of an 
electromagnetic field. 
Then, Eq.(\ref{307}) becomes 
\begin{equation}
 \exp ( \delta s \, \gamma ^{\mu } \:\! {\partial }_{\mu } ) \Psi = 
 \exp ( - {\rm i } {\rm \, \delta } \;\!\! s \, {\rm e } \gamma ^{\mu } 
 \! {A }_{\mu } ) \Psi . \label{318}
\end{equation}
We consider Eq.(\ref{318}) to be the fundamental equation for a spread 
electron in an electromagnetic field. 
\subsection{\label{subsec:level32}Derivation of the wave equation}
In Eq.(\ref{318}), we substitute 
\begin{eqnarray}
 \begin{aligned}
 \left\{
 \begin{array}{ll}
 X \; {\equiv } \;\; \delta s \, \gamma ^{\mu } \:\! 
 {\partial }_{\mu } , \label{319} \\
 \;\! Y \; {\equiv } \;\; {\rm i } \, \delta s \, 
 {\rm e } \gamma ^{\mu } \! {A }_{\mu } . 
 \end{array} \right. 
 \end{aligned}
\end{eqnarray}
Then, 
\begin{equation}
 \exp ( X ) \Psi = \exp (- Y ) \Psi . \label{320}
\end{equation}
It follows that 
\begin{equation}
 \{ {\exp ( X ) \exp ( Y ) } \} \exp ( - Y ) \Psi 
 = \exp (- Y ) \Psi . \label{321} 
\end{equation}
According to the Campbell-Hausdorff formula: 
\begin{eqnarray}
 \begin{aligned}
 \exp ( X ) \exp ( Y ) = \exp ( Z ) , \ \ \mbox{where} \hspace{65mm} \\ 
 Z = X + Y 
 + \frac{1 }{2 } [ \, X , \, Y \, ] 
 + \frac{1 }{12 } {\{} \, [ \, [ \, X , \, Y \, ], \, Y \, ] 
 - [ \, [ \, X , \, Y \,] , \, X \, ] \, {\}} 
 + \cdots ,  \label{322} 
 \end{aligned}
\end{eqnarray}
equation (\ref{321}) becomes 
\begin{equation}
 \{ \exp ( Z ) - {\rm I } \} \exp (- Y ) \Psi= 0 . 
 \label{323} 
\end{equation}
We can expand $ \{ \exp ( Z ) - {\rm I } \} $ 
into the infinite product of a sine function as follows: 
\begin{eqnarray}
 \begin{aligned}
 \exp ( Z ) - {\rm I } \hspace{103.5mm} \\
 = \exp \Big( \frac{ Z }{2 } \Big) 
 \Big\{ \exp \Big( \frac{ Z }{2 } \Big) 
 - \exp \Bigl( \! - \frac{ Z }{2 } \Big) \Bigr\} \hspace{61.5mm} \\ 
 = - 2 \, {\rm i } \exp \Big( \frac{ Z }{2 } \Big) \: 
 {\sin } \Big( \frac{ {\rm i } Z }{2 } \Big) \hspace{83mm} \\
 = \exp \Big( \frac{ Z }{2 } \Big) Z 
 {\prod_{ n = 1 }^{\infty } } 
 \Bigl\{ \Big( {\rm I } - \frac{ {\rm i } Z }{ 2 \:\! n \pi } \Big) 
 \exp \Big( \frac{ {\rm i } Z }{ 2 \:\! n \pi } \Big) \Bigr\} 
 \Bigl\{ \Big( {\rm I } + \frac{ {\rm i } Z }{ 2 \:\! n \pi } \Big) 
 \exp \Big( \! - \frac{ {\rm i } Z }{ 2 \:\! n \pi } \Big) \Bigr\} . 
 \hspace{0mm} \label{324} 
 \end{aligned}
\end{eqnarray}
Thus, the equation that $ \Psi $ should satisfy is 
\begin{equation}
 ( \:\! {\rm i } Z - 2 \:\! n \pi ) \exp ( - Y ) \Psi = 0 
 \;;\;\; n = 0, \pm 1, \pm 2, \cdots . \label{325} 
\end{equation}
Using the expansion 
\begin{equation}
 \phi \, \exp ( {\rm i } \, \omega )
 = \exp ( {\rm i } \, \omega ) \{ \, { \phi 
 + {\rm i } \, [ \, \phi , \, \omega \, ] 
 - \frac{1 }{2 } 
 [ \, [ \, \phi , \, \omega \, ] , \, \omega \, ] 
 + \cdots } \, \} , 
 \label{326} 
\end{equation}
and substituting 
\begin{eqnarray}
 \begin{aligned}
 \left\{
 \begin{array}{ll}
 \phi \; {\equiv } \; {\rm i } Z - 2 \:\! n \pi , 
 \label{327} \\
 \omega \; {\equiv } \; {\rm i } Y ,
 \end{array} \right. 
 \end{aligned}
\end{eqnarray}
we can obtain the following wave equation: 
\begin{equation}
 \Bigl( {\rm i } \gamma ^{\mu } \:\! \partial _{\mu } - {\rm e } 
 \gamma ^{\mu } A_{\mu } + V ( \delta s ) 
 - \frac{2 n \pi }{\delta s } \Bigr) \Psi = 0 , \label{328} 
\end{equation}
where $ \delta s $ is determined as the Compton wavelength 
$ \lambda _{c } $($ = 2 \pi / m $) of the electron 
and $ n $ is selected as $ 1 $ such that the mass term of the electron 
is specified correctly. 

In this case, wave equation (\ref{328}) agrees with 
the Dirac equation except for the existence of $ V (\delta s ) $. 
Here, $ V (\delta s ) $ is represented as a power series of $ \delta s $ 
as follows: 
\begin{equation}
 V ( \delta s ) = V _{1 } + V _{2 } 
 + \cdots , \hspace{58mm} \label{329} 
\end{equation}
\vspace{-7mm}
\addtocounter{equation}{-1}
\begin{subequations}
\begin{eqnarray}
 V _{1 } \! \! & \equiv & \! \! + \frac{1 }{2 } \delta s \, 
 {\rm e } \, [ \, \gamma ^{\mu } \:\! \partial _{\mu }, \, \gamma ^{\nu } 
 A_{\nu } \, ] , \label{329a} \\
 V _{2 } \! \! & \equiv & \! \! - \frac{1 }{12 } {\rm i } \, 
 \delta s^{2 } \, 
 {\rm e } \, [ \, [ \gamma ^{\mu } \:\! \partial _{\mu }, \, \gamma ^{\nu } 
 A_{\nu } \, ], \, \gamma ^{\xi } 
 ( {\rm i } \partial _{\xi } + {\rm e } A_{\xi }) \, ] . \hspace{6mm} 
 \label{329b} 
\end{eqnarray}
\end{subequations}

In addition, if $ \delta s $ is the infinitesimal, 
$ V ( \delta s ) \to 0 $ and the mass term corresponds to the infinite 
bare electron mass. 
Then, the usual Dirac equation is reproduced when the mass term is 
renormalized. 
Therefore, the Dirac equation is inherently relativistic invariant. 
However, we assume another standpoint because $ V ( \delta s ) $ 
has the physical significance as shown below. 
\section{\label{sec:level4}Calculation of the Anomalous Magnetic Moment}
\hspace*{0.6cm}In this section, by considering $ V _n $ as the nth order 
correction of the Dirac equation, 
we evaluate the corrections in magnetic moment using the 
Foldy-Wouthuysen transformation \cite{foldy} 
(FW transformation). 
The result obtained is in good agreement with the QED calculation. 
\subsection{\label{subsec:level41}FW transformation of the Dirac equation}
We begin with the FW transformation of the Dirac equation: 
\begin{equation}
 ( {\rm i } \gamma ^{\mu } \:\! \partial _{\mu } - {\rm e } \gamma ^{\mu } 
 A_{\mu } - m ) \Psi = 0 . \label{401}
\end{equation}
Note that $ \beta \equiv \gamma ^{0 }, \ \beta ^{2 } = 1, \ 
\boldsymbol{\alpha } \equiv \beta \boldsymbol{\gamma } $, 
$ \partial_{\mu } \equiv ( \partial / \partial t, \boldsymbol{\nabla } ) $, 
and $ A_{\mu } \equiv (\phi , - \! \boldsymbol{A } ) $. 
Thus, multiplying the left-hand side of Eq.(\ref{401}) with the $ \beta $ 
matrix, we obtain a time-independent Dirac Hamiltonian: 
\begin{equation}
 {\mathcal{H } } = \beta m + \varepsilon + o \hspace{45mm} \label{402} 
\end{equation}
\vspace*{-9mm}
\addtocounter{equation}{-1}
\begin{subequations}
\begin{eqnarray}
 \mbox{with the even operator} \ \ \varepsilon \! & = & \! {\rm e } \phi 
 \label{402a} \hspace{10mm} \\
 \mbox{and the odd operator} \ \ o \! & = & \! \boldsymbol{\alpha } \, 
 {\dotprod } \, \boldsymbol{\pi } ; \label{402b} \hspace{10mm} \\ 
 \boldsymbol{\pi } \equiv \boldsymbol{p } 
 - {\rm e } \boldsymbol{A } 
 \! \! & \equiv & \! \! - {\rm i } \boldsymbol{\nabla } 
 - {\rm e } \boldsymbol{A } . \nonumber 
\end{eqnarray}
\end{subequations}
Performing the FW transformation eliminates the odd operator from 
$ {\mathcal{H } } $: 
\begin{equation}
 {\mathcal{H^{\textsc{FW}} } } \simeq \beta m + \varepsilon + \frac{1 }{2 m } 
\beta o^{2 } 
 - \frac{1 }{8 m^{2 } } [ \, o , \, [ \, o , \, \varepsilon \, ] \, ] . 
 \label{403} 
\end{equation}
Using the identity 
\begin{equation}
 ( \boldsymbol{\alpha } {\dotprod } \, \boldsymbol{a } ) 
 ( \boldsymbol{\alpha } {\dotprod } \, \boldsymbol{b } ) 
 = \boldsymbol{a } \, {\dotprod } \, \boldsymbol{b } 
 + {\rm i } \boldsymbol{\sigma } {\dotprod } 
 ( \boldsymbol{a } \! \boldsymbol{\times } \! 
 \boldsymbol{b } ) , \label{404} 
\end{equation}
where $ \boldsymbol{a } $ and $ \boldsymbol{b } $ are arbitrary 
vectors and $ \boldsymbol{\sigma } $ denotes the $ 4 \times 4 $ Dirac spin 
matrix, we can obtain the explicit form of Eq.(\ref{403}) as follows: 
\begin{eqnarray}
 \begin{aligned}
 {\mathcal{H^{\textsc{FW}} } } \simeq \frac{1 }{2 m } \beta 
 \boldsymbol{\pi }^{2 } + {\rm e } \phi + \beta m  \hspace{46mm} \\
 - \frac{ {\rm e } }{2 m } \frac{ \textit{\textsf{g}} }{2} \beta 
 \boldsymbol{\sigma } 
 {\dotprod } \boldsymbol{B } - \frac{1 }{2 } \frac{ {\rm e } }{2 m } 
 \frac{1 }{m } \boldsymbol{\sigma } {\dotprod } 
 ( \boldsymbol{E \! \times \! \pi } )
 + \frac{ {\rm e } }{8 m^2 } \boldsymbol{\nabla }^2 \phi , \hspace{0mm} \\
 \mbox{where} \quad 
 \boldsymbol{B } = \boldsymbol{\nabla \! \times \! A } , \ \boldsymbol{E } 
 = - \boldsymbol{\nabla } \phi , \label{405} 
 \end{aligned}
\end{eqnarray}
and the gyromagnetic ratio $ \textit{\textsf{g}} $ of an electron described 
by the Dirac equation is only $ 2 $. 
\subsection{\label{subsec:level42}Effect of $ V _{1 } $}
We evaluate the alteration in $ {\mathcal{H^{\textsc{FW}} } } $ 
by adding 
$ V _{1 } $ to the Dirac equation. 
$ V _{1 } $ is rewritten as follows: 
\begin{eqnarray}
 \begin{aligned}
 V _{1 } = + \kappa \frac{ {\rm e } }{4 m } 
 \left\{ \sigma ^{\mu \nu } 
 F_{\mu \nu } - 2 \sigma ^{\mu \nu } ( A_{\mu } \partial _{\nu } 
 - A_{\nu } \partial _{\mu } ) \right\} , \hspace{18mm} \\
 \mbox{where} \quad 
 \kappa \equiv - {\rm i } \, \delta s \, m = - 2 \pi {\rm i } , 
 \hspace{36mm} \\ 
 \sigma ^{\mu \nu } 
 \equiv \frac{ {\rm i } }{2 } [ \, \gamma ^{\mu }, \, \gamma ^{\nu } \, ] , \ 
 \mbox{and} \ \ 
 F_{\mu \nu } \equiv \frac{\partial A_{\nu } } 
 {\partial x^{\mu } } - \frac{\partial A_{\mu } }{\partial x^{\nu } } . 
 \label{406} 
 \end{aligned}
\end{eqnarray}
This expression contains the Pauli term 
$ \kappa \left( {\rm e } / 4 m \right) \sigma ^{\mu \nu } F_{\mu \nu } $; 
therefore, it appears to be related to the magnetic moment. 
Further, $ V _{1 } $ can also be expressed as follows: 
\begin{equation}
 V _{1 } = - \kappa \frac{ {\rm e } }{2 m } 
 \Bigl\{ ( \boldsymbol{\sigma } {\dotprod } \boldsymbol{B } 
 - {\rm i } \boldsymbol{\alpha } {\dotprod } \boldsymbol{E } ) 
 - 2 \Bigl\{ \boldsymbol{\sigma } {\dotprod }
 ( \boldsymbol{ A \! \times \! \nabla } ) 
 - {\rm i } \boldsymbol{\alpha } {\dotprod } \Bigl( \phi \boldsymbol{\nabla } 
 + \boldsymbol{A } \frac{\partial}{\partial t} 
 \Bigr) \! \Bigr\} \! \Bigr\} , \label{407} 
\end{equation}
where alterations in (\ref{402a}) and (\ref{402b}) are 
\addtocounter{equation}{-1}
\begin{subequations}
\begin{eqnarray}
 \delta \varepsilon \! \! & = & \! \! + \kappa \frac{ {\rm e } }{2 m } 
 \beta \boldsymbol{\sigma } {\dotprod } \left( \boldsymbol{B } 
 - 2 \boldsymbol{ A \! \times \! \nabla } \right) \label{408a} \\
 \mbox{and} \ \ \delta o \! \! & = & \! \! - \kappa \frac{ {\rm e } }{2 m } 
 \beta \, {\rm i } \boldsymbol{\alpha } {\dotprod } 
 \Bigl\{ \boldsymbol{E } - 2 \Bigl( \phi \boldsymbol{\nabla } 
 + \boldsymbol{A } \frac{\partial}{\partial t} 
 \Bigr) \! \Bigr\} . \label{408b} 
\end{eqnarray}
\end{subequations}
We then calculate the alterations in $ {\mathcal{H^{\textsc{FW}} } } $ 
by using identity (\ref{404}),
\begin{eqnarray}
 \begin{aligned}
 \delta \Bigl\{ \frac{1 }{2 m } \beta o^{2 } \Bigr\} \simeq
 \frac{1 }{2 m } \beta ( \, o \, \delta o + \delta o \, o \, ) 
 \hspace{46.5mm} \\ 
 \quad = + \kappa \frac{ {\rm e } }{4 m^2 } {\rm i } \, 
 \boldsymbol{\sigma } {\dotprod } 
 \left( \boldsymbol{ \nabla \! \times \! E } \right) 
 + \kappa \frac{ {\rm e } }{4 m^2 } \left( \boldsymbol{\nabla } {\dotprod } 
 \boldsymbol{E } + \boldsymbol{E } {\dotprod } \boldsymbol{\nabla } \right) 
 \label{409} \hspace{0mm} \\
 \qquad - \kappa \frac{ {\rm e } }{2 m } \beta \, \boldsymbol{\sigma } 
 {\dotprod } 
 \left( \boldsymbol{B } - 2 \boldsymbol{ A \! \times \! \nabla } \right) 
 \! \Bigl\{ \frac{\beta }{m } \Bigl( {\rm i } \frac{\partial}{\partial t} 
 - {\rm e } \phi \Bigr) \! \Bigr\} \hspace{2.5mm} \\
 - \kappa \frac{ {\rm e }^2 }{2 m^2 } \boldsymbol{\sigma } {\dotprod } 
 \left( \boldsymbol{ A \! \times \! E } \right) \hspace{45.5mm} 
 \end{aligned}
\end{eqnarray}
\vspace*{-4mm}
and 

\begin{eqnarray}
 \begin{aligned}
 \delta \Bigl\{ - \frac{1 }{8 m^{2 } } 
 [ \, o, \, [ \, o, \, \varepsilon \, ] \, ] \Bigr\} \hspace{67mm} \\
 \simeq - \frac{1 }{8 m^{2 } } {\{} [ \, \delta o, \, 
 [ \, o, \, \varepsilon \, ] \, ] 
 \! + \! [ \, o , \, [ \, \delta o , \, \varepsilon \, ] \, ] 
 \! + \! [ \, o , \, [ \, o , \, \delta \varepsilon \, ] \, ] {\}} 
 \hspace{0mm} \\ 
 \simeq - \kappa \frac{ {\rm e }^2 }{4 m^2 } {\rm i } \, 
 \boldsymbol{A } {\dotprod } \boldsymbol{E } \Bigl\{ \frac{\beta }{m } 
 \Bigl( {\rm i } \frac{\partial}{\partial t} - {\rm e } \phi \Bigr) \! 
 \Bigr\} . \hspace{29mm} \label{410}  
 \end{aligned}
\end{eqnarray}
Consequently, the alteration in $ {\mathcal{H^{\textsc{FW}} } } $ 
due to $ V _{1 } $ can be expressed as follows: 
\begin{eqnarray}
 \begin{aligned}
 \delta {\mathcal{H^{\textsc{FW}} } } \simeq \delta 
 \varepsilon 
 + \delta \Bigl\{ \frac{1 }{2 m } \beta o^{2 } \Bigr\} 
 + \delta \Bigl\{ - \frac{1 }{8 m^{2 } } 
 [ \, o , \, [ \, o , \, \varepsilon \, ] \, ] \Bigr\} \hspace{17.5mm} \\
 \simeq  
 + \kappa \frac{ {\rm e } }{4 m^2 } {\rm i } \, \boldsymbol{\sigma } 
 {\dotprod } \left( \boldsymbol{ \nabla \! \times \! E } \right) 
 + \kappa \frac{ {\rm e } }{4 m^2 } \left( \boldsymbol{\nabla } {\dotprod } 
 \boldsymbol{E } + \boldsymbol{E } {\dotprod } \boldsymbol{\nabla } \right) 
 \hspace{11mm} \\
 - \kappa \frac{ {\rm e } }{2 m } \beta \boldsymbol{\sigma } 
 {\dotprod } 
 \left( \boldsymbol{B } - 2 \boldsymbol{ A \! \times \! \nabla } \right) \! 
 \Bigl\{ \frac{\beta }{m } \Bigl( {\rm i } \frac{\partial}{\partial t} 
 - {\rm e } \phi - \beta m \Bigr) \! \Bigr\} \hspace{3.5mm} \label{411} \\
 - \kappa \frac{ {\rm e }^2 }{2 m^2 } \boldsymbol{\sigma } 
 {\dotprod } \left( \boldsymbol{ A \! \times \! E } \right) 
 - \kappa \frac{ {\rm e }^2 }{4 m^2 } {\rm i } \, \boldsymbol{A } {\dotprod } 
 \boldsymbol{E } \Bigl\{ \frac{\beta }{m } 
 \Bigl( {\rm i } \frac{\partial}{\partial t} 
 - {\rm e } \phi \Bigr) \! \Bigr\} . 
 \end{aligned}
\end{eqnarray}
We now assume the following conditions: 
\begin{itemize}
 \item The external electromagnetic field is sufficiently small and static. 
 \item The kinetic energy is sufficiently smaller than the rest energy 
of the electron. 
\end{itemize}
Then, in Eq.(\ref{411}), $ \boldsymbol{ \nabla \! \times \! E } = 0 $ 
and $ {\rm i } ( \partial / \partial t ) - {\rm e } \phi \simeq \beta m $. 
In addition, since the scalar potential $ \phi $ is time-independent, 
it commutes with $ {\mathcal{H^{\textsc{FW}} } } $, i.e., 
\begin{equation}
 0 = \frac{d \phi }{dt } = {\rm i } \, [{\mathcal{H^{\textsc{FW}} } } , 
 \phi \, ] \, 
 \simeq \beta \frac{ {\rm i } }{2 m } \left( \, p^{2 } \phi - \phi \, 
 p^{2 } \, \right)
 = \beta \frac{ {\rm i } }{2 m } \left( \, \boldsymbol{\nabla } 
 {\dotprod } \boldsymbol{E } 
 + \boldsymbol{E } {\dotprod } \boldsymbol{\nabla } \, \right) . 
 \label{412} 
\end{equation}
Therefore, Eq.(\ref{411}) becomes 
\begin{equation}
 \delta {\mathcal{H^{\textsc{FW}} } } \simeq 
 - \kappa \frac{ {\rm e }^2 }{2 m^2 } \boldsymbol{\sigma } {\dotprod } 
 \left( \boldsymbol{ A \! \times \! E } \right) 
 - \kappa \frac{ {\rm e }^2 }{4 m^2 } {\rm i } \, \boldsymbol{A } {\dotprod } 
 \boldsymbol{E } . \label{413} 
\end{equation}
Since $ \boldsymbol{ A \! \times \! E } $ and 
$ \boldsymbol{A } {\dotprod } \boldsymbol{E } $ are sufficiently small, 
$ \delta {\mathcal{H^{\textsc{FW}} } } $ can be neglected 
as compared to $ {\mathcal{H^{\textsc{FW}} } } $. 
Hence, Eq.(\ref{328}) roughly agrees with the Dirac equation. 
\subsection{\label{subsec:level43}Self-energy influence}
Here, we assume that there exist no external electric charges. 
According to classical electromagnetics, the electron obtains 
its self-energy in the form of electrostatic energy. 
In addition, the electrostatic energy $ {\rm e } \phi $ in the Dirac 
Hamiltonian differs from $ m $ by a factor of the $ \beta $ matrix: 
\begin{equation}
 {\mathcal{H } } \simeq \beta m + {\rm e } \phi 
 = \beta \left( m + \beta {\rm e } \phi \right) . \label{414} 
\end{equation}
Hence, the self-energy can be defined as 
$ \delta m \equiv \beta {\rm e } \phi $ such that $ \delta m $ may behave 
as a part of $ m $. 

We now evaluate the alteration $ \delta {\mathcal{H^{\textsc{FW}} } } $ 
by taking the self-energy into consideration. 
Here, the electric field generated by the rest electron is 
$ \boldsymbol{E } = - ( \phi / r^2 ) \boldsymbol{r } $, and the vector 
potential for a constant magnetic field is 
$ \boldsymbol{A }= (1/2) \boldsymbol{B \! \times \! r } $, where 
$ \boldsymbol{r } $ is the position vector from the charge. 
Then, the second term in (\ref{413}) becomes 
zero, since 
$
 \boldsymbol{A } {\dotprod } \boldsymbol{E } 
 = - (1/2) ( \phi / r^2 ) \left( \boldsymbol{B \! \times \! r } \right) 
 {\dotprod } \, \boldsymbol{r } = 0 . 
$
On the other hand, 
$
 \left( \boldsymbol{B \! \times \! r } \right) \! \boldsymbol{\times r } 
 = - \boldsymbol{B } \left( \boldsymbol{r } \, {\dotprod } \, 
 \boldsymbol{r } \right) + \boldsymbol{r } 
 \left( \boldsymbol{B } \, {\dotprod } \, \boldsymbol{r } \right) 
 = - \boldsymbol{B } \, r^2 
$, 
since the mean value of $ \boldsymbol{B } \, {\dotprod } \, \boldsymbol{r } $ 
becomes zero due to the spherical symmetry of $ \boldsymbol{r } $. 
Therefore, the first term in (\ref{413}) becomes 
\begin{eqnarray}
 \begin{aligned}
 - \kappa \frac{ {\rm e }^2 }{2 m^2 } \boldsymbol{\sigma } {\dotprod } 
 \left( \boldsymbol{A \! \times \! E } \right)
 = - \frac{1 }{2 } \Bigl( \kappa \frac{ \beta {\rm e } \phi }{m } \Bigr) 
 \frac{ {\rm e } }{2 m } \, \frac{ \beta \, \boldsymbol{\sigma } {\dotprod } 
 \{ - \! \left( \boldsymbol{B \! \times \! r } \right) \! 
 \boldsymbol{\times r } \} }{r^2 } \\
 = - \frac{1 }{2 } \Bigl( \kappa \frac{\delta m }{m } \Bigr) 
 \frac{ {\rm e } }{2 m } \beta \, \boldsymbol{\sigma } {\dotprod } 
 \boldsymbol{B } . \hspace{25mm} \label{415} 
 \end{aligned}
\end{eqnarray}
It is observed that (\ref{415}) gives a correction in 
the magnetic moment that is proportional to $ \delta m $. 
\subsection{\label{subsec:level44}Self-energy estimation}
In this study, we assumed that an electron has a time-like size 
$ \delta s $ as the world length. 
We then interpret $ 0.5 \, \delta s \left( = \pi / m \right) $ as the 
four-dimensional radius $ r_{0 } $ of the electron, which is the same degree 
of Zitterbewegung amplitude being expected from the Dirac equation. 

The classical calculation provides 
a good approximation of the self-energy since the quantum effects 
are insignificant when $ r_{0 } > 1/m $, even if these effects take part in. 
In order to estimate the self-energy of an electron that is spread in four 
dimensions, we extend the definition of the electric field, as shown below, 
by applying Gauss's law to the surface of a four-dimensional sphere. 
The area of the four-dimensional sphere is $ 2 \pi^2 r^3 $ and the electric 
charge on the sphere is $ \rm e $ multiplied by $ r_0 $, hence, the 
four-dimensional electric field $ \widetilde{ \boldsymbol{E} } $ shall be 
defined as 
\begin{equation}
 \widetilde{ \boldsymbol{E} } = \frac{ r_0 {\rm e }}{ 2 \pi^2 \epsilon _0 r^3 } 
 \frac{ \boldsymbol{r} }{r}, \ \ \ \label{416} 
\end{equation}
where $ \epsilon _0 $ denotes the dielectric constant of vacuum. 

Thus, the four-dimensional self-energy $ \widetilde{\delta m} $ of an electron 
can be estimated by an analogy with the classical electrostatic energy as 
\begin{eqnarray}
 r_0 \delta m = \widetilde{\delta m} 
 = \frac{\epsilon _0}{2} \int { \widetilde{ \boldsymbol{E} } }^2 
 {\rm d }^4 r. \hspace{0mm} \label{417} 
\end{eqnarray}
Therefore, 
\begin{eqnarray}
 \begin{aligned}
 \delta m = \frac{\epsilon _0}{2 r_0 } \int 
 \widetilde{ \boldsymbol{E} }^2 {\rm d }^4 r 
 = \frac{\epsilon _0}{2 r_0 } \int _{r _0}^{\infty} 
 \Bigl( \frac{ r_0 {\rm e }}{ 2 \pi^2 \epsilon _0 r^3 } \Bigr)^{ 2 } 
 2 \pi^2 r^3 {\rm i } {\rm d } r \\
 = {\rm i } \frac{ {\rm e }^2} 
 { 8 \pi^2 \epsilon _0 r _0 } , \hspace{63.5mm} \label{418} 
 \end{aligned}
\end{eqnarray}
where spatial integration is performed with respect to the imaginary 
radius $ {\rm i } r $, since an electron has a time-like spread. 
Thus, the self-energy of an electron with a time-like spread becomes an 
imaginary number and it is not observable. 

Nevertheless, by substituting (\ref{418}) into (\ref{415}), 
we can obtain the first-order correction of the magnetic moment: 
\begin{equation}
 - \frac{1 }{2 } \Bigl( \kappa \frac{\delta m }{m } \Bigr) 
 \frac{ {\rm e } }{2 m } 
 \beta \, \boldsymbol{\sigma } {\dotprod } \boldsymbol{B } 
 = - \frac{1 }{2 } \Bigl( \frac{\alpha }{ \pi } \Bigr) \frac{ {\rm e } }{2 m } 
 \beta \, \boldsymbol{\sigma } {\dotprod } \boldsymbol{B } , \label{419} 
\end{equation}
where $ \alpha \left( \equiv {\rm e }^2 / 4 \pi \epsilon _0 \right) $ denotes 
the fine structure constant. 
Accordingly, the correction in the gyromagnetic ratio can be expressed as 
\begin{equation}
 \frac{\textit{\textsf{g}} - 2 }{2 } = \frac{1 }{2 } 
\left(\frac{\alpha }{\pi } \right) ; \label{420} 
\end{equation}
this expression agrees with the calculation by J.~Schwinger (1948) 
\cite{schwinger}. 
\subsection{\label{subsec:level45}Higher-order correction in the magnetic 
moment}
Finally, we calculate the $ \alpha^2 $-order correction in the magnetic 
moment. 
We use the symbol $ \delta^{(2 ) } $ to denote second-order variations 
and omit calculations that do not directly contribute to the magnetic moment. 
We now expand (\ref{329b}) as follows: 
\begin{eqnarray}
 V _{2 } = \kappa ^2 \frac{ {\rm e } }{12 m^2 } {\rm i } \, 
 [ \, [ \gamma ^{\mu } \:\! \partial _{\mu }, \, \gamma ^{\nu } 
 A_{\nu } \, ], \, 
 \gamma ^{\xi } \left( {\rm i } \partial _{\xi } 
 + {\rm e } A_{\xi } \right) \, ] , \ \ \ \label{421} 
\end{eqnarray}
\vspace*{-7mm}
where 
\begin{eqnarray}
 & & [ \, [ \gamma ^{\mu } \:\! \partial _{\mu }, \, \gamma ^{\nu } 
 A_{\nu } \, ], \, 
 \gamma ^{\xi } \left( {\rm i } \partial _{\xi } + {\rm e } A_{\xi } \right) 
 \, ] \nonumber \\
 & & = \nonumber \\
 (a) \hspace{5mm} & & +2 \, \gamma ^{\xi } \gamma ^{\nu } 
 \left( \partial _{\xi } A_{\nu } \right) \left( {\rm i } \gamma ^{\mu } 
 \partial _{\mu } - {\rm e } \gamma ^{\mu } A_{\mu } \right) \nonumber \\
 (b) \hspace{5mm} & & + 4 A_{\nu } \gamma ^{\mu } 
 \left( {\rm i } \partial _{\mu } 
 - {\rm e } A_{\mu } \right) \partial ^{\, \nu } \nonumber \\
 (c) \hspace{5mm} & & - 4 {\rm i } \, \gamma ^{\nu } A_{\nu } \, 
 \gamma ^{\xi } \partial _{\xi } \, 
 \gamma ^{\mu } \partial _{\mu } \nonumber \\
 (d) \hspace{5mm} & & + 4 {\rm e } \, \gamma ^{\nu } A_{\nu } \, 
 \gamma ^{\xi } A_{\xi } \, \gamma ^{\mu } \partial _{\mu } \label{422} \\
 (e) \hspace{5mm} & & - 2 {\rm i } \, \gamma ^{\mu } \left( \partial _{\nu } 
 A_{\mu } \right) \partial ^{\, \nu } \nonumber \\
 (f) \hspace{5mm} & & - 4 {\rm e } \, \gamma ^{\nu } \left( \partial _{\mu } 
 A_{\nu } \right) A^{\mu } \nonumber \\
 (g) \hspace{5mm} & & + 6 {\rm e } \, \gamma ^{\mu } \left( \partial _{\mu } 
 A_{\nu } \right) A^{\nu } . \nonumber 
\end{eqnarray}
In the following, we evaluate each term in (\ref{422}). 
Note that 
$ {\rm e } \gamma ^{0 } A_{0 } = \beta {\rm e } \phi = \delta m $. 
Thus, we have 
\begin{eqnarray}
 \begin{aligned}
 \mbox{(a):} \hspace{4mm} + 2 \, \gamma ^{\xi } \gamma ^{\nu } 
 \left( \partial _{\xi } A_{\nu } \right) 
 \left( {\rm i } \gamma ^{\mu } \partial _{\mu } - {\rm e } \gamma ^{\mu } 
 A_{\mu } \right) \hspace{61mm} \\
 \simeq + 2 {\rm i } \, \boldsymbol{\sigma } {\dotprod } 
 \boldsymbol{B } \left( {\rm i } \gamma ^{\mu } \partial _{\mu } 
 - {\rm e } \gamma ^{\mu } A_{\mu } \right)
 + 2 \, \boldsymbol{\alpha } {\dotprod } \boldsymbol{E } 
 \left( {\rm i } \gamma ^{0 } \partial _{0 } 
 - {\rm e } \gamma ^{0 } A_{0 } \right) \hspace{28.0mm} \label{423} \\
 \simeq + 2 {\rm i } \, \boldsymbol{\sigma } {\dotprod } 
 \boldsymbol{B } \, m - 2 {\rm i } \, \beta \, \boldsymbol{\alpha } 
 {\dotprod } \boldsymbol{E } \, 
 \partial _{0 } - 2 \, \boldsymbol{\alpha } {\dotprod } \boldsymbol{E } \, 
 \delta m . \hspace{50.5mm}
 \end{aligned}
\end{eqnarray}
The first term in (\ref{423}) results in the following alteration in 
$ {\mathcal{H^{\textsc{FW}} } } $: 
\begin{equation}
 \delta^{(2 ) } \varepsilon _{a1 } 
 = - \beta \, \kappa ^2 \frac{ {\rm e } }{12 m^2 } {\rm i } 
 \left( + 2 {\rm i } \, \boldsymbol{\sigma } {\dotprod } \boldsymbol{B } \, 
 m \right)
 = + \frac{1 }{3 } \kappa ^2 
 \frac{ {\rm e } }{2 m } \beta \, \boldsymbol{\sigma } {\dotprod } 
 \boldsymbol{B } . \label{424} 
\end{equation}
Although (\ref{424}) contributes to the magnetic moment, 
it will be counterbalanced by another correction term 
that will be calculated later in (\ref{431}). 
The second term in (\ref{423}) is unrelated to the magnetic moment 
since the $ \boldsymbol{\sigma } $ matrix does not appear 
in the result of the FW transformation. 
The third term in (\ref{423}) contributes to the magnetic moment; 
hence, it will be evaluated below together with the (g) term. 

The terms (b), (c), and (d) might contribute to the magnetic moment 
through the variations in $ \boldsymbol{p } $ and $ \boldsymbol{A } $ 
of (\ref{402b}). 
It should be noted that $ {\rm i } \gamma ^{0 } \partial _{0 } 
\simeq m + \delta m $ 
and $ {\rm e } \gamma ^{0 } A_{0 } = \delta m $. Thus, we have 
\begin{eqnarray}
 \begin{aligned}
 \mbox{(b):} \hspace{4mm} + 4 A_{\nu } \gamma ^{\mu } 
 \left( {\rm i } \partial _{\mu } - {\rm e } A_{\mu } \right) 
 \partial ^{\, \nu } \hspace{74.5mm} \\
 \simeq + \frac{4 {\rm i } }{ {\rm e } } \sum _{k} {\rm e } 
 \gamma ^{0 } A_{0 } \, \gamma ^{k } 
 \left( {\rm i } \partial _{k } + {\rm e } A_{k } \right) {\rm i } 
 \gamma ^{0 } \partial _{0 }
 - \frac{4 {\rm i } }{ {\rm e } } \, {\rm e } 
 \gamma ^{0 } A_{0 } \, \gamma ^{0 } 
 \left( {\rm i } \partial _{0 } - {\rm e } A_{0 } \right) {\rm i } 
 \gamma ^{0 } \partial _{0 } \label{425} \hspace{0mm} \\
 \simeq - \frac{4 {\rm i } }{ {\rm e } } \sum _{k} \gamma ^{k } 
 \left( - {\rm i } \partial _{k } - {\rm e } A_{k } \right) 
 \delta m \left( m + \delta m \right)
 - \frac{4 {\rm i } }{ {\rm e } } \, \delta m \, m 
 \left( m + \delta m \right) , \hspace{9.0mm} 
 \end{aligned}
\end{eqnarray}
\begin{eqnarray}
 \begin{aligned}
 \mbox{(c):} \hspace{4mm} - 4 {\rm i } \, \gamma ^{\nu } A_{\nu } \, 
 \gamma ^{\xi } \partial _{\xi } \, \gamma ^{\mu } \partial _{\mu } 
 \hspace{82.5mm} \\
 \simeq - \frac{4 {\rm i } }{ {\rm e } } \sum _{k} {\rm e } 
 \gamma ^{k } A_{k } \, {\rm i } \gamma ^{0 } \partial _{0 } \, 
 {\rm i } \gamma ^{0 } \partial _{0 }
 + \frac{4 {\rm i } }{ {\rm e } } \, {\rm e } \gamma ^{0 } A_{0 } \, {\rm i } 
 \gamma ^{0 } \partial _{0 } \, {\rm i } \gamma ^{0 } \partial _{0 } 
 \hspace{32.5mm} \\
 - \frac{4 {\rm i } }{ {\rm e } } \sum _{k} {\rm e } \gamma ^{0 } 
 A_{0 } \, ( \gamma ^{0 } \gamma ^{k } + \gamma ^{k } \gamma ^{0 } ) \, 
 \partial _{0 } \, \partial _{k } \hspace{58.0mm} \label{426} \\
 \simeq + \frac{4 {\rm i } }{ {\rm e } } \sum _{k} \gamma ^{k } 
 \left( - {\rm e } A_{k } \right) \left( m + \delta m \right) ^2 
 + \frac{4 {\rm i } }{ {\rm e } } \, \delta m \left( m + \delta m \right) ^2 
 , \hspace{28.0mm}
 \end{aligned}
\end{eqnarray}
\begin{eqnarray}
 \begin{aligned}
 \mbox{(d):} \hspace{4mm} + 4 {\rm e } \, \gamma ^{\nu } A_{\nu } \, 
 \gamma ^{\xi } A_{\xi } \, \gamma ^{\mu } \partial _{\mu } 
 \hspace{82.0mm} \\
 \simeq - \frac{4 {\rm i } }{ {\rm e } } \sum _{k} {\rm e } 
 \gamma ^{0 } A_{0 } \, {\rm e } \gamma ^{0 } A_{0 } \, {\rm i } \gamma ^{k } 
 \partial _{k }
 - \frac{4 {\rm i } }{ {\rm e } } \, {\rm e } \gamma ^{0 } 
 A_{0 } \, {\rm e } \gamma ^{0 } A_{0 } \, {\rm i } \gamma ^{0 } 
 \partial _{0 } \hspace{30.0mm} \\
 + \frac{4 {\rm i } }{ {\rm e } } \sum _{k} \, 
 ( \gamma ^{0 } \gamma ^{k } + \gamma ^{k } \gamma ^{0 } ) \, 
 {\rm e } ^2 A_{0 } A_{k } \, {\rm i } \gamma ^{0 } \partial _{0 } 
 \label{427} \hspace{55.0mm} \\
 \simeq + \frac{4 {\rm i } }{ {\rm e } } \sum _{k} \gamma ^{k } 
 \left( - {\rm i } \partial _{k } \right) \delta m ^2
 - \frac{4 {\rm i } }{ {\rm e } } \, \delta m ^2 \left( m + \delta m \right) 
 . \hspace{42.0mm}
 \end{aligned}
\end{eqnarray}
After collecting the terms (b), (c), and (d), we get the following two terms: 
\begin{equation}
 + \frac{4 {\rm i } }{ {\rm e } } \sum _{k} \! \gamma ^{k } \! 
 \left( - {\rm e } A_{k } \right) m^2 \! 
 + \frac{4 {\rm i } }{ {\rm e } } \sum _{k} \! \gamma ^{k } \! 
 \left( {\rm i } \partial _{k } - {\rm e } A_{k } \right) m \, \delta m . 
 \label{428} 
\end{equation}
These terms result in the following alterations in $ {\mathcal{H } } $: 
\begin{eqnarray}
 \begin{aligned}
 \delta^{(2 ) } o_{bcd} 
 \simeq - \beta \, \kappa ^2 \frac{ {\rm e } }{12 m^2 } 
 {\rm i } \, \frac{4 {\rm i } }{ {\rm e } } \sum _{k} \! \gamma ^{k } \! 
 \left( - {\rm e } A_{k } \right) m^2 \hspace{16.0mm} \\
 - \beta \, \kappa ^2 \frac{ {\rm e } }{12 m^2 } {\rm i } \, 
 \frac{4 {\rm i } }{ {\rm e } } \sum _{k} \! \gamma ^{k } \! 
 \left( {\rm i } \partial _{k } - {\rm e } A_{k } \right) m \, \delta m 
 \label{429} \hspace{4.5mm} \\
 = + \frac{1 }{3 } \kappa ^2 \! 
 \left( - {\rm e } \, \boldsymbol{\alpha } {\dotprod } \boldsymbol{A } \right) 
 + \frac{1 }{3 } \kappa ^2 \frac{\delta m }{m } 
 \, \boldsymbol{\alpha } {\dotprod } 
 \left( - \boldsymbol{p } - {\rm e } \boldsymbol{A } \right) . 
 \end{aligned}
\end{eqnarray}
Then, vector potential $ \boldsymbol{A } $ in (\ref{402b}) 
is corrected by the first term in (\ref{429}) as 
\begin{equation}
 \boldsymbol{A } \to \Bigl( 1 + \frac{1 }{3 } \kappa ^2 \Bigr) 
 \boldsymbol{A } . \label{430} 
\end{equation}
Accordingly, the magnetic moment in (\ref{405}) is corrected as 
\begin{equation}
 - \frac{ {\rm e } }{2 m } \beta \, \boldsymbol{\sigma } {\dotprod } 
 \boldsymbol{B } \to - \Bigl( 1 + \frac{1 }{3 } \kappa ^2 \Bigr) 
 \frac{ {\rm e } }{2 m } \beta \, \boldsymbol{\sigma } {\dotprod } 
 \boldsymbol{B } . \label{431} 
\end{equation}
The variation in (\ref{431}) is counterbalanced by 
$ \delta^{(2 ) } \varepsilon_{a1 } $, which was previously calculated. 
In the second term of (\ref{429}), corrections in the magnetic moment 
due to variations in $ \boldsymbol{p } $ and $ \boldsymbol{A } $ cancel 
each other out in the result of the FW transformation. 

The (e) and (f) terms can be neglected under the conditions assumed in 
Subsection \ref{subsec:level42}. 
\begin{eqnarray}
 \begin{aligned}
 \mbox{(g):} \hspace{4mm} + 6 {\rm e } \, \gamma ^{\mu } 
 \left( \partial _{\mu } A_{\nu } \right) A^{\nu } \hspace{85.0mm} \\
 \simeq + 6 \sum _{k} \gamma ^{0 } \gamma ^{k } \! 
 \left( - \partial _{k } A_{0 } \right) {\rm e } \gamma ^{0 } A_{0 } 
 \simeq + 6 \, \boldsymbol{\alpha } {\dotprod } 
 \boldsymbol{E } \, \delta m . \hspace{42.0mm} \label{432} 
 \end{aligned}
\end{eqnarray}
This term contributes to the magnetic moment as well as the third term 
of (a). 

In any case, we obtain the alteration in $ {\mathcal{H } } $ related to the 
magnetic moment by adding (g) and the third term of (a) as follows: 
\begin{equation}
 \delta^{(2 ) } o_{ a3 + g } = - \beta \, \kappa ^2 
 \frac{ {\rm e } }{12 m^2 } {\rm i } \left( - 2 \, \boldsymbol{\alpha } 
 {\dotprod } \boldsymbol{E } \, \delta m 
 + 6 \, \boldsymbol{\alpha } {\dotprod } \boldsymbol{E } \, 
 \delta m \right)
 = - \frac{1 }{3 } \kappa ^2 
 \frac{ \delta m }{m } \frac{ {\rm e } }{m } {\rm i } \, \beta \, 
 \boldsymbol{\alpha } {\dotprod } \boldsymbol{E } . \label{433} 
\end{equation}
This term results in the following alteration in 
$ {\mathcal{H^{\textsc{FW}} } } $: 
\begin{eqnarray}
 \begin{aligned}
 \delta^{(2 ) } \Bigl\{ \frac{1 }{2 m } \beta \, 
 o \, ^2 \Bigr\}_{ a3 + g }
 \simeq \frac{1 }{2 m } \beta \{ \, o \, \delta^{(2 ) } 
 o_{ a3 + g } \, + \delta^{(2 ) } o_{ a3 + g } \, o \, \} \hspace{28mm} \\
 \simeq - \frac{1 }{3 } \kappa ^2 \frac{ \delta m }{m } 
 \frac{ {\rm e } }{2 m^2 } {\rm i } \, \beta 
 \{ \left( - {\rm e } \, \boldsymbol{\alpha } {\dotprod } 
 \boldsymbol{A } 
 \right) \left( \beta \, \boldsymbol{\alpha } {\dotprod } \boldsymbol{E } 
 \right) + \left( \beta \, \boldsymbol{\alpha } {\dotprod } \boldsymbol{E } 
 \right) \left( - {\rm e } \, \boldsymbol{\alpha } {\dotprod } 
 \boldsymbol{A } \right) \} \label{434} \\
 = + \frac{1 }{3 } \Bigl( \kappa \frac{ \delta m }{m } \Bigr) 
 \Bigl\{ \kappa \frac{ {\rm e }^2 }{m^2 } \, \boldsymbol{\sigma } {\dotprod } 
 \left( \boldsymbol{ A \! \times \! E } \right) \Bigr\}
 = 
 + \frac{1 }{3 } \left( \frac{\alpha }{\pi } \right)^2 \! \! 
 \frac{ {\rm e } }{2 m } \beta \, \boldsymbol{\sigma } {\dotprod } 
 \boldsymbol{B } . \hspace{8mm} 
 \end{aligned}
\end{eqnarray}
Consequently, the correction in the gyromagnetic ratio up to the order of 
$ \alpha^2 $ becomes 
\begin{equation}
 \frac{\textit{\textsf{g}} - 2 }{2 } = \frac{1 }{2 } \left( \frac{\alpha } 
 {\pi } \right) 
 - \frac{1 }{3 } \left( \frac{\alpha }{\pi } \right)^2 , \hspace{3mm} \label{435} 
\end{equation}
whereas the corresponding correction calculated in QED is \ 
\cite{sommerfield, petermann1} 
\begin{equation}
 \frac{\textit{\textsf{g}}_{qed} - 2 }{2 } = \frac{1 }{2 } 
 \left( \frac{\alpha }{\pi } \right) 
 - 0.3285 \left( \frac{\alpha }{\pi } \right)^2 . \label{436} 
\end{equation} 
Both these results agree within the error margin of the order of $ \alpha^3 $. 
\subsection{\label{subsec:level46}Vacuum polarization effect}
In the previous sections, the effect of vacuum polarization is not taken 
into account. 
Hence, we consider that the marginal error between (\ref{435}) 
and (\ref{436}) can be attributed to vacuum polarization due to the 
electron pair creation. 
The size of the error can be approximated by the following expression: 
\begin{equation}
 \frac{\textit{\textsf{g}}_{qed} }{2} - \frac{ \, \textit{\textsf{g}} \, }{2} 
 = \Bigl\{ - 0.3285 + \frac{1}{3} \, \Bigr\} \! 
 \left( \frac{\alpha }{\pi } \right)^2 
 \simeq + \frac{1 }{12 } \! \left( \frac{\alpha }{4 \pi } \right)^2 
 \! \! . \ \label{437} 
\end{equation}
Accordingly, we assume that the alteration in 
$ {\mathcal{H^{\textsc{FW}} } } $ 
due to the electron pair creation is given by the following formula: 
\begin{equation}
 \delta {\mathcal{H^{\textsc{FW}} }_{{\rm e }^+ {\rm e }^- } } 
 \simeq - \frac{1 }{12 } \! 
 \left( \frac{\alpha }{4 \pi } \right)^2 \! \! \frac{ {\rm e } }{2 m } \beta 
 \, \boldsymbol{\sigma } {\dotprod } \boldsymbol{B } . \label{438} 
\end{equation}

Then, the correction in the gyromagnetic ratio for the $ \alpha^2 $ order 
is recalculated as 
\begin{equation}
 - \Bigl\{ + \frac{1}{3} - \frac{1 }{12 } 
 \Bigl( \frac{ \, 1 \, }{4} \Bigr)^2 \, \Bigr\} 
 \! \left( \frac{\alpha }{\pi } \right)^2 \! 
 \simeq - 0.3281 \left( \frac{\alpha }{\pi } \right)^2 . \label{439} 
\end{equation}
In fact, this value is almost in agreement with that of the QED calculation 
and differs from the experimental value by only 
$ 1.5 \left( \alpha /\pi \right)^3 $ \cite{wesley}. 

The above assumption is found to be appropriate by estimating the 
muon magnetic moment in which the influence of vacuum polarization is 
more significant. 
The vacuum polarization due to the muon pair creation yields a correction 
similar to that observed in Eq.(\ref{438}): 
\begin{equation}
 \delta {\mathcal{H^{\textsc{FW}} }_{\mu^+ \mu^- } } \simeq - \frac{1 }{12 } 
 \! \left( \frac{\alpha }{4 \pi } \right)^2 \! \! 
 \frac{ {\rm e } }{2 m_\mu } \beta \, \boldsymbol{\sigma } {\dotprod } 
 \boldsymbol{B } , \label{440} 
\end{equation}
where $ m_\mu $ denotes the muon mass. 
In addition, the effect of electron pair creation exists. 
With regard to the muon magnetic moment, we simply assume that 
the effect of electron pair creation 
is the same as that given by Eq.(\ref{438}): 
\begin{equation}
 \delta {\mathcal{H^{\textsc{FW}} }_{{\rm e }^+ {\rm e }^- } } 
 \simeq - \frac{1 }{12 } \! 
 \left( \frac{\alpha }{4 \pi } \right)^2 \! \! \frac{ {\rm e } }{2 m } \beta 
 \, \boldsymbol{\sigma } {\dotprod } \boldsymbol{B }
 = - \frac{1 }{12 } 
 \! \left( \frac{\alpha }{4 \pi } \right)^2 \! \! 
 \Bigl( \frac{ \, m_\mu }{m } \Bigr) 
 \frac{ {\rm e } }{2 m_\mu } \beta \, \boldsymbol{\sigma } {\dotprod } 
 \boldsymbol{B } , \label{441} 
\end{equation}
where the mass ratio $ m_\mu / m $ is around $ 206.8 $. 

Then, the correction in the muon gyromagnetic ratio for the $ \alpha^2 $ order 
is obtained by adding (\ref{440}), (\ref{441}), and (\ref{434}) 
for the muon mass: 
\begin{eqnarray}
 - \Bigl\{ + \frac{1}{3} - \frac{1 }{12 } \Bigl( \frac{ \, 1 \, }{4} \Bigr)^2 
 \! \Bigl( 1 + \! \frac{ \, m_\mu }{m } \Bigr) \Bigr\} \! 
 \left( \frac{\alpha }{\pi } \right)^2 \! \simeq 0.75 \left( \frac{\alpha }
 {\pi } \right)^2 . \label{442} 
\end{eqnarray}
This value agrees with that obtained by the QED calculation 
\cite{suura, petermann2}. 

Therefore, the second-order correction (\ref{435}) is also considered to be 
an appropriate result when the effect of vacuum polarization is not taken 
into account. 
\section{\label{sec:level5}Many-Coordinate Systems Interpretation}
\hspace*{0.6cm}We assumed that an electron has an inherent relativistic 
symmetry. 
In other words, all the inertial coordinate systems in Minkowski space 
are symmetric and superpositioned from a viewpoint of a free electron. 
In this context, the measurement process is also explained as symmetry 
breaking caused by the observation from a specific inertial coordinate system. 
This results in a nonlocal stochastic process because for an electron, the 
measurement implies an unpredictable selection of a specific coordinate 
system in which the observation is performed. 

For example, quantum entanglement, i.e., the EPR correlation \cite{epr} 
of pair particles with opposite helicity, is prepared by the superposition 
of a right- and a left-handed coordinate system, which 
correspond to either of the eigenstates of helicity. 
The observation of helicity in one particle concurrently fixes the state 
of another particle through the selection of either of the coordinate systems. 

The many-coordinate systems interpretation presented here 
is similar to the many-worlds interpretation of Everett 
\cite{everett} \textit{et al}. 
They propose the existence of many worlds corresponding to the 
superpositioned eigenstates. 
However, it is not the worlds but the coordinate systems that will branch 
because of the observation. 
Special relativity guarantees that all the coordinate systems that may 
branch exist in a Minkowski space. 
In addition, we consider that the material particle in classical mechanics 
is a substance in which relativistic symmetry is almost lost 
due to the coupling of a large number of elementary particles. 
\section{\label{sec:level6}Conclusion}
\hspace*{0.6cm}In this study, we assumed that the quantum behavior of an 
electron lies in its relativistic symmetry. 
Based on this idea, we derived a Lorentz invariant equation for the spread 
electron and demonstrated the validity of the equation by calculating the 
anomalous magnetic moment without renormalization. 
In addition, based on the same idea, we consistently explained the measurement 
process in quantum theory. 
The calculation method in the present paper is not practical since the 
electromagnetic interaction is not the minimal one and is not gauge invariant. 
However, an inherent relativistic symmetry holds true also for the 
dimensionless electron described by the unrenormalized Dirac equation. 
We conclude that the foundations of quantum mechanics will be understood 
only in relation to relativistic symmetry; 
this is the only manner in which the foundations of both theories can be 
bridged within a conventional Minkowski space. 
\newpage
\end{document}